\documentclass[conference]{IEEEtran}
\usepackage{amsmath,amssymb,epsfig}
\usepackage{amsmath,amssymb,epsfig,cite,url}
\usepackage{subfigure}
\usepackage{multirow}
\usepackage{algpseudocode}
\usepackage{algorithm}
\usepackage{cite,url}
\newtheorem{theorem}{\underline{Theorem}}[section]
\newcommand{\mv}[1]{\mbox{\boldmath{$ #1 $}}}
\psfull
\begin{document}
\title{Channel Knowledge Map for Environment-\\Aware Communications: EM Algorithm \\for Map Construction}
\author{\IEEEauthorblockN{Kun~Li\IEEEauthorrefmark{1}, Peiming Li\IEEEauthorrefmark{1}\IEEEauthorrefmark{3}, Yong~Zeng\IEEEauthorrefmark{1}\IEEEauthorrefmark{2},~\text{and}~Jie Xu\IEEEauthorrefmark{4}}
\IEEEauthorblockA{\IEEEauthorrefmark{1}
National Mobile Communications Research Laboratory, Southeast University, Nanjing 210096, China}
\IEEEauthorblockA{\IEEEauthorrefmark{2}
Purple Mountain Laboratories, Nanjing 211111, China}
\IEEEauthorblockA{\IEEEauthorrefmark{3}
School of Information Engineering, Guangdong University of Technology, Guangzhou 510006, China}
\IEEEauthorblockA{\IEEEauthorrefmark{4}
FNii and SSE, The Chinese University of Hong Kong, Shenzhen, Shenzhen 518172, China}
E-mail:
kun\_lee@seu.edu.cn, peiminglee@outlook.com, yong\_zeng@seu.edu.cn, xujie@cuhk.edu.cn\vspace{-0pt}
}
\maketitle

\newcommand\blfootnote[1]{
\begingroup
\renewcommand\thefootnote{}\footnote{#1}
\addtocounter{footnote}{-1}
\endgroup
}
\begin{abstract}
Channel knowledge map (CKM) is an emerging technique to enable environment-aware wireless communications, in which databases with location-specific channel knowledge are used to facilitate or even obviate real-time channel state information acquisition. One fundamental problem for CKM-enabled communication is how to efficiently construct the CKM based on finite measurement data points at limited user locations. Towards this end, this paper proposes a novel map construction method based on the \emph{expectation maximization} (EM) algorithm, by utilizing the available measurement data, jointly with the expert knowledge of well-established statistic channel models. The key idea is to partition the available data points into different groups, where each group shares the same modelling parameter values to be determined. We show that determining the modelling parameter values can be formulated as a maximum likelihood estimation problem with latent variables, which is then efficiently solved by the classic EM algorithm. Compared to the pure data-driven methods such as the nearest neighbor based interpolation, the proposed method is more efficient since only a small number of modelling parameters need to be determined and stored. Furthermore, the proposed method is extended for constructing a specific type of CKM, namely, the channel gain map (CGM), where closed-form expressions are derived for the \emph{E-step} and \emph{M-step} of the EM algorithm. Numerical results are provided to show the effectiveness of the proposed map construction method as compared to the benchmark curve fitting method with one single model.
\end{abstract}
\section{Introduction}
Channel knowledge map (CKM) is an emerging technique towards environment-aware wireless communications \cite{CKM}, which provides location-specific (rather than the coarse site-specific) channel knowledge associated with potential transmitter-receiver pairs by, e.g., storing them in databases. Compared to conventional environment-ignorant communication, CKM-enabled environment-aware communication is expected to facilitate or even obviate real-time channel state information (CSI) acquisition, which makes it especially appealing for future communication systems with large spatial dimensions \cite{Array} and prohibitive channel training overhead.

In fact, the attempts to use site-specific databases in wireless communications have been pursued in prior works based on, e.g., 3D city or terrain map \cite{R1}, radio environment map \cite{TVBand1,R5}, and TV white space map \cite{TVBand2}. However, these designs require storing accurate physical environment maps and implementing computation-expensive algorithms, such as ray tracing algorithms, which are costly in terms of both storage and computation. Furthermore, the TV white space map and radio environment map were mainly used for cognitive radio systems \cite{REM}, for which the obtained maps critically depend on the status/activities of the primary transmitters, such as the spectrum, power, and antenna pattern being used. By contrast, CKM aims to provide location-specific knowledge that directly reflects the intrinsic channel characteristics, regardless of the transmitter or receiver activities \cite{CKM}. This makes it possible to design communication systems with light or even without real-time channel training \cite{Beam-TrainFree}. Some specific instances of CKM include channel gain map (CGM) \cite{R2}, channel path map (CPM) \cite{Beam-TrainFree}, and beam index map (BIM) \cite{Beam-TrainFree,R3}. CKM-enabled communications have been recently studied in various applications, such as training-free subband selection for device-to-device (D2D) communications \cite{CKM}, beam alignment for millimeter wave (mmWave) massive MIMO \cite{Beam-TrainFree}, and trajectory design for cellular-connected UAV \cite{R4}.

One fundamental problem for CKM-enabled environment-aware communication is how to efficiently construct the CKM based on finite measurement data points at limited user locations. The most straightforward approach for map construction is interpolation-based methods, such as the inverse distance weighted (IDW) \cite{IDW}, nearest neighbours (NN), splines \cite{Spline}, and Kriging \cite{Kriging1,Spatial}. The interpolation-based methods, however, generally require large measurement data for accurate map construction. In addition, such pure data-driven methods ignored the well-established stochastic or geometric based channel models developed over the past few decades \cite{Channelmodels}, thus usually requiring huge storage capacity for map maintenance. To utilize both measurement data and expert knowledge, one straightforward method is parametric curve fitting, where the best modelling parameters of the selected channel models are determined based on the measurement data. However, such a naive curve fitting method would lead to poor map quality, since the number of tunable modelling parameters is typically very small, while the environment is usually too complex to be accurately characterized by one single model.

To overcome the above drawbacks, in this paper, we propose a novel CKM construction method based on the well-established \emph{expectation maximization} (EM) algorithm \cite{EM}. Notice that  the naive curve fitting method is not able to accurately predict the complex channel environment knowledge, as only one common channel model is used. By contrast, it is observed that different sub-areas of the site may experience different radio propagation environment, which may be modelled with different sets of modelling parameters. Based on this observation, we propose to partition the available measurement data into different modelling groups, where each group shares the same modelling parameter values that are to be determined. We show that the determination of modelling parameter values corresponds to a maximum likelihood estimation problem with latent variables. Although this is a challenging non-convex optimization problem, various existing algorithms, such as the classic EM algorithm \cite{EM}, have been proposed to find its efficient solutions. In particular, we propose a generic EM-based algorithm to solve the considered CKM construction problem, which consists of two steps, namely the Expectation step (\emph{E-step}) and the Maximization step (\emph{M-step}). Furthermore, we also consider the special case for constructing a CGM, for which we extend the EM-based algorithm to find the optimized solution, by deriving the closed-form expressions for the \emph{E-step} and \emph{M-step}, respectively. Finally, extensive numerical results are provided to verify the effectiveness of the proposed EM-based algorithm for CKM construction.
\section{System Model}
\begin{figure}[t]
\centerline{\includegraphics[width=6cm]{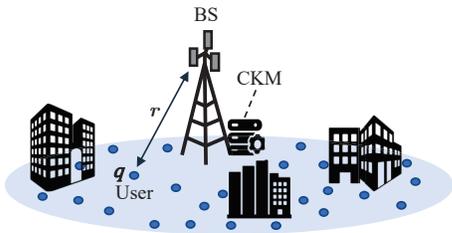}}
\caption{CKM-enabled environment-aware wireless communications.}
\label{senerio}
\end{figure}

As shown in Fig. \ref{senerio}, we consider a wireless communication system in a specific site, with a stationary base station (BS) and mobile users, whose potential locations $\boldsymbol{q}$ are denoted by the set ${\mathcal Q}$. For any given ${\mv q} \in {\mathcal Q}$, our objective is to predict the interested location-specific channel knowledge, which is denoted as $\boldsymbol{r}$, as accurate as possible before real-time channel training is applied. Note that the channel knowledge $\boldsymbol{r}$ can be any useful information related to the wireless channel, such as the channel gain, shadowing, angle of arrival/departure (AoA/AoD), or even the channel impulse response. To this end, a BS-to-any (B2X) CKM $\mathcal W$ is constructed, which provides mapping from location $\boldsymbol{q}$ to the corresponding channel knowledge $\boldsymbol{r}$, i.e., $ \mathcal W: \boldsymbol{q} \in \mathcal Q \to \boldsymbol{r}$.

One immediate method for channel knowledge prediction based on user locations is to utilize the well-established channel models, which are usually given stochastically with certain modelling parameter vector $\boldsymbol{\theta}$, denoted as $p(\boldsymbol{r}|\boldsymbol{q},\boldsymbol{\theta})$. Specifically, $p(\boldsymbol{r}|\boldsymbol{q},\boldsymbol{\theta})$ gives the probability density function (PDF) of the channel knowledge $\boldsymbol{r}$ for users located at $\boldsymbol{q}$, parameterized by $\boldsymbol{\theta}$. As a concrete example, for the specific B2X CGM, $\boldsymbol{r}$ corresponds to the channel gain in dB, which is denoted by the real number $r$. Without loss of generality, we may assume that the BS is located at the origin. Therefore, according to the classic path loss model, we have
\begin{align}
r=\beta+10\alpha \log_{10}{\|\boldsymbol{q}\|}+S,\label{eq1}
\end{align}
where $\alpha$ denotes the path loss exponent, $\beta$ denotes the path loss intercept, $S\sim {\cal N}(0,\sigma^2)$ captures the log-normal shadowing with variance $\sigma^2$, and $\|\cdot\|$ denotes the Euclidean norm. As such, the modelling parameter vector for channel gains is $\boldsymbol{\theta}=[\alpha, \beta, \sigma^2]$, and the conditional PDF of the channel knowledge $r$ given user location $\boldsymbol{q}$ is
\begin{align}
&p(r|\boldsymbol{q},\boldsymbol{\theta})
=N(r|\beta+10\alpha\log_{10}{\|\boldsymbol{q}\|},\sigma^2)\notag\\
&~~~~~~=\frac{1}{\sqrt{2\pi\sigma^2}}\exp{\left( {-(r-\beta-10\alpha\log_{10}{\|\boldsymbol{q}\|})^2}/{{2\sigma^2}}\right)},\label{eq2}
\end{align}
where $N(r|\mu, \sigma^2)$ denotes the PDF of a Gaussian random variable $r$ with mean $\mu$ and variance $\sigma^2$. A direct channel knowledge prediction based on stochastic channel models would lead to poor accuracy, since such models characterize channels only on the average sense, by ignoring the site-specific or even location-specific propagation environment. For instance, with the channel model in \eqref{eq1}, no matter how good the modelling parameter $\boldsymbol{\theta}$ is chosen, the predicted channel gain is symmetric around the BS, which is far from reality when the actual environment is taken into account.

On the other hand, site-specific or location-specific radio propagation environment information can be learned if on-site measurement data are available. Let $\boldsymbol{X} \in \mathbb{R}^{D \times N}$ denote the set of measurement data points. Each column of $\boldsymbol{X}$, denoted as $\boldsymbol{x}_n\in \mathbb{R}^{D\times 1}, n \in {\cal N}\triangleq\{1,...,N\}$, corresponds to one data point, which includes the measured channel knowledge $\boldsymbol{r}_n$ at the corresponding location $\boldsymbol{q}_n$. Therefore, we have $\boldsymbol{x}_n=[\boldsymbol{q}_n^T, \boldsymbol{r}_n^T]^{T}$, with the superscript $T$ denoting the transpose. Note that the number of available measurement data points $N$ is usually limited. As a result, to construct a complete CKM for all potential user locations, one needs to infer the channel knowledge of those unmeasured locations based on $\boldsymbol{X}$. However, the pure data-driven methods, such as the interpolation-based methods, usually require large data and high storage capacity. On the other hand, the simple parameteric curve fitting based method, which uses one single model to fit all data points, would lead to poor accuracy due to the limited degrees of freedom associated with the few modelling parameters.

To overcome the above issues, we propose a map construction method based on multi-component or mixed channel models. Specifically, with the basic PDF $p(\boldsymbol{r}|\boldsymbol{q},\boldsymbol{\theta})$, we consider a mixed channel model with a total of $K$ components, corresponding to $K$ sets of modelling parameters, denoted as $\boldsymbol{\theta}_k, k\in{\cal K}\triangleq \{1,...,K\}$. With a slight abuse of notation, let $\boldsymbol{\theta}=\{\boldsymbol{\theta}_1, ....,\boldsymbol{\theta}_K\}$. Then for the given measurement data $\boldsymbol{X}$, our objective is to find the set of modelling parameters $\boldsymbol{\theta}$ based on certain criterion. A commonly used criterion is the maximum likelihood estimation, which corresponds to finding parameters ${\mv \theta}$ for maximizing the likelihood as follows.
\begin{align}
\textrm{(P1):} \max_{\boldsymbol{\theta}} p(\boldsymbol{X}|\boldsymbol{\theta}),\notag
\end{align}
where $p(\boldsymbol{X}|\boldsymbol{\theta})$ denotes the likelihood function with respect to the modelling parameters $\boldsymbol{\theta}$.
\section{EM Algorithm for CKM Construction}
The key challenge of solving \textrm{(P1)} lies in finding the explicit expression for the likelihood function $p(\boldsymbol{X}|\boldsymbol{\theta})$. This is difficult since for each data point $\boldsymbol{x}_n \in \boldsymbol{X}$, it is unknown which of the $K$ modelling components it should be associated with. As a consequence, \textrm{(P1)} corresponds to the maximum likelihood estimation problem with latent variables, which, fortunately, has been extensively studied and can be efficiently solved by various algorithms, such as the classic EM algorithm\cite{EM}. To this end, for each data point $\boldsymbol{x}_n, n\in{\cal N}$, we introduce a vector of latent variables $\boldsymbol{z}_n\in \mathbb{R}^{K\times1}$ to indicate the association of data point $\boldsymbol{x}_n$ with the $K$ modelling components, whose $k$-th element $z_{nk}$ is a binary random variable with $z_{nk}\in\{0,1\}, \forall k\in{\cal K}$. Here, $z_{nk}=1$ means that $\boldsymbol{x}_n$ is best explained by the $k$-th modelling component with parameter $\boldsymbol{\theta}_k$. It follows that $\sum_{k\in \mathcal K} z_{nk} = 1$, as each data point $\boldsymbol{x}_n$ is only associated with the best modelling component. Furthermore, the distribution of $\boldsymbol{z}_n$ is specified by the mixing coefficients $\pi_k$, i.e.,
\begin{align}
p(z_{nk}=1)=\pi_k, k\in \cal K,
\end{align}
where $0\leq \pi_k\leq 1$ and $\sum\nolimits_{k\in{\cal K}}\pi_k=1$. As such, for any given modelling parameters $\boldsymbol{\theta}$, the conditional PDF of $\boldsymbol{x}_n$ given its associated latent variable $\boldsymbol{z}_n$ is\cite{EM}
\begin{align}
p(\boldsymbol{x}_n|\boldsymbol{z}_n,\boldsymbol{\theta})=
p(\boldsymbol{x}_n|\boldsymbol{\theta}_{k^{*}}) =
\prod\nolimits_{k\in{\cal K}} [p(\boldsymbol{x}_n|\boldsymbol{\theta}_k)]^{z_{nk}},\label{eq5}
\end{align}
where $k^*$ defined such that $z_{nk^*}=1$. Note that the last equality in \eqref{eq5} holds since $z_{nk}\in \{0,1\}, \forall k\in \cal K$.
Similarly, $p(\boldsymbol{z}_n|\boldsymbol{\theta})$ can be expressed as
\begin{align}
&p(\boldsymbol{z}_n|\boldsymbol{\theta})= \prod\nolimits_{k\in{\cal K}} \pi_k^{z_{nk}}.\notag
\end{align}
As a result, the joint PDF of $\boldsymbol{x}_n$ and $\boldsymbol{z}_n$ can be expressed as
\begin{align}
p(\boldsymbol{x}_n,\boldsymbol{z}_n|\boldsymbol{\theta})
=p(\boldsymbol{x}_n|\boldsymbol{z}_n,\boldsymbol{\theta})p(\boldsymbol{z}_n|\boldsymbol{\theta})=\prod\limits_{k\in{\cal K}}[\pi_kp(\boldsymbol{x}_n|\boldsymbol{\theta}_k)]^{z_{nk}}.\label{eq7}
\end{align}
Let $\boldsymbol{Z} \in \mathbb{R}^{K \times N}$ include all the $N$ vectors $\boldsymbol{z}_n$ of latent variables, $n \in \cal N$. Since different data points are independent, we have $p(\boldsymbol{X},\boldsymbol{Z}|\boldsymbol{\theta}) =\prod_{n \in {\cal N}} p(\boldsymbol{x}_n,\boldsymbol{z}_n|\boldsymbol{\theta})$. Therefore, the likelihood function $p(\boldsymbol{X}|\boldsymbol{\theta})$ is obtained by marginalizing $p(\boldsymbol{X},\boldsymbol{Z}|\boldsymbol{\theta})$ over $\boldsymbol{Z}$, and \textrm{(P1)} can be equivalently written as
\begin{align}
\textrm{(P2):} \max\limits_{\boldsymbol{\theta}} \sum\nolimits_{\boldsymbol{Z}} p(\boldsymbol{X},\boldsymbol{Z}|\boldsymbol{\theta}).\notag
\end{align}
Note that $p({\mv X}|{\mv \theta})$ and $p(\boldsymbol{X},\boldsymbol{Z}|\boldsymbol{\theta})$ are referred to as the likelihood functions of the {\it incomplete-data} and {\it complete-data}\cite{EM}, respectively. \textrm{(P2)} is the maximum likelihood estimation problem with the latent variable $\boldsymbol{Z}$, which can be efficiently solved by the classic EM algorithm\cite{EM}. The EM algorithm is an iterative algorithm with two basic steps, i.e., the \emph{E-step} and the \emph{M-step}, which are summarized in Algorithm 1 \cite[Section 9.3]{EM}. It is shown in \cite[Section 9.4]{EM} that in the EM algorithm, each step will lead to an increased (or at least non-deceased) {\it{complete-data}} log likelihood. Therefore, the convergence of the EM algorithm is guaranteed \cite{EM}.
\begin{algorithm}[ht]
\renewcommand{\algorithmicrequire}{\textbf{Input:}}
\renewcommand{\algorithmicensure}{\textbf{Output:}}
\small
  \caption{General EM Algorithm \cite[Section 9.3]{EM}.}
  \begin{algorithmic}[1]
    \Require Given the joint distribution $p(\boldsymbol{X},\boldsymbol{Z}|\boldsymbol{\theta})$ governed by the set of parameters $\boldsymbol{\theta}$;
    \State {\bf Initialization:} Choose initial parameters $\boldsymbol{\theta}^\text{old}$ and mixing coefficients $\pi_k^\text{old}$, $\forall k \in \cal K$;
    \State {\bf Repeat:}
        \begin{itemize}
            \item[1)] {\bf E-step}: Evaluate the posterior distribution of the latent variables $p(\boldsymbol{Z}|\boldsymbol{X},\boldsymbol{\theta}^\text{old})$ and the {\it{responsibilities}} $\{\gamma_{nk}=\mathbb{E}[z_{nk}]\}$;
            \item[2)] {\bf M-step}: Update $\boldsymbol{\theta}^\text{new}=\arg\max\limits_{\boldsymbol{\theta}}Q(\boldsymbol{\theta},\boldsymbol{\theta}^\text{old})
                \triangleq \sum_{\boldsymbol{Z}}p(\boldsymbol{Z}|\boldsymbol{X},\boldsymbol{\theta}^\text{old})\ln p(\boldsymbol{X},\boldsymbol{Z}|\boldsymbol{\theta})$, and $\pi^{\text{new}}_k={N_k}/{N}$, with $N_k=\sum_{n\in{\cal N}}\gamma_{nk}$;
            \item[3)] $\boldsymbol{\theta}^{\text{old}}$ $\leftarrow$ $\boldsymbol{\theta}^{\text{new}}$, $\pi_k^{\text{old}}$ $\leftarrow$ $\pi_k^{\text{new}}$, $\forall k \in \cal K$;
        \end{itemize}
    \State {\bf Until} convergence or a maximum number of iterations is reached;
    \Ensure Set of modelling parameters $\boldsymbol{\theta}^{\text{new}}$, and {\it{responsibilities}} $\{\gamma_{nk}\}$.
  \end{algorithmic}
\end{algorithm}
\subsection{General Algorithm for EM-based CKM Construction}\label{A}
In the following, the \emph{E-step} and the \emph{M-step} are developed for our considered CKM construction problem \textrm{(P2)}.
\subsubsection{E-step}
As given in Algorithm 1, the \emph{E-step} of the EM algorithm is to evaluate the posterior distribution of the latent variable $p(\boldsymbol{Z}|\boldsymbol{X},\boldsymbol{\theta}^\text{old})$, with given modelling parameter $\boldsymbol{\theta}^\text{old}$. According to equation (9.75) of\cite{EM}, $p(\boldsymbol{Z}|\boldsymbol{X},\boldsymbol{\theta}^\text{old})$ can be factorized as
\begin{align}
p(\boldsymbol{Z}|\boldsymbol{X},\boldsymbol{\theta}^\text{old})
&=\prod\nolimits_{n\in \cal N} p(\boldsymbol{z}_n|\boldsymbol{x}_n,\boldsymbol{\theta}^\text{old}).\label{eq9}
\end{align}
Furthermore, with the Bayesian theorem, we have
\begin{align}
p(\boldsymbol{z}_n|\boldsymbol{x}_n,\boldsymbol{\theta}^\text{old})=\frac{p(\boldsymbol{z}_n|\boldsymbol{\theta}^\text{old})p(\boldsymbol{x}_n|\boldsymbol{z}_n,\boldsymbol{\theta}^\text{old})}{\sum_{\boldsymbol{z}'_n}p(\boldsymbol{z}'_n|\boldsymbol{\theta}^\text{old})p(\boldsymbol{x}_n|\boldsymbol{z}'_n,\boldsymbol{\theta}^\text{old})}.\label{eq10}
\end{align}

Let $\gamma_{nk}$ denote the {\it{responsibility}} corresponding to component $k\in{\cal K}$ to explain the measurement data $\boldsymbol{x}_n$, which can be obtained as the expected value of the indicator variable $z_{nk}$ with the posterior distribution in \eqref{eq10}, i.e.,
\begin{align}
\gamma_{nk}&\triangleq \mathbb{E}[z_{nk}]=p(z_{nk}=1|\boldsymbol{x}_n,\boldsymbol{\theta}^\text{old})\notag\\
&={\pi^\text{old}_k p(\boldsymbol{x}_n|\boldsymbol{\theta}^\text{old}_k)}/{\sum\nolimits_{j\in \cal K} \pi^\text{old}_j p(\boldsymbol{x}_n|\boldsymbol{\theta}^\text{old}_j)}\notag\\
&={\pi^\text{old}_k p(\boldsymbol{r}_n|\boldsymbol{q}_n,\boldsymbol{\theta}^\text{old}_k)}/{\sum\nolimits_{j\in \cal K} \pi^\text{old}_j p(\boldsymbol{r}_n|\boldsymbol{q}_n,\boldsymbol{\theta}^\text{old}_j)}, \label{B1}
\end{align}
where the second equality holds since $z_{nk}$ is binary. Furthermore, the last equality of \eqref{B1} follows by noting that $\boldsymbol{x}_n=[\boldsymbol{q}_n^T, \boldsymbol{r}_n^T]^T$, so that
\begin{align}
p(\boldsymbol{x}_n|\boldsymbol{\theta}_k)=p(\boldsymbol{q}_n,\boldsymbol{r}_n|\boldsymbol{\theta}_k)=p(\boldsymbol{r}_n|\boldsymbol{q}_n,\boldsymbol{\theta}_k), \label{eq8}
\end{align}
where $p(\boldsymbol{r}_n|\boldsymbol{q}_n,\boldsymbol{\theta}_k)$ corresponds to the selected stochastic channel model, such as \eqref{eq2} for CGM. As a result, with the \emph{E-step}, the posterior probabilities $p(\boldsymbol{Z}|\boldsymbol{X},\boldsymbol{\theta}^\text{old})$ is obtained based on \eqref{eq9} and \eqref{eq10}, and the {\it{responsibilities}} $\{\gamma_{nk}\}$ is obtained based on \eqref{B1}.
\subsubsection{M-step}
Based on the posterior probabilities in \eqref{eq9} and the {\it responsibilities} in \eqref{B1} obtained in the \emph{E-step}, the modelling parameter $\boldsymbol{\theta}$ and mixing coefficients $\pi_k$ are updated in the \emph{M-step}, by maximizing the expectation of the log-likelihood of the {\it{complete-data}}, with the expectation taken with respect to $p(\boldsymbol{Z}|\boldsymbol{X},\boldsymbol{\theta}^\text{old})$, which is given by  \cite{EM}
\begin{align}
Q(\boldsymbol{\theta},\boldsymbol{\theta}^\text{old})=\sum\nolimits_{\boldsymbol{Z}}p(\boldsymbol{Z}|\boldsymbol{X},\boldsymbol{\theta}^\text{old})\ln p(\boldsymbol{X},\boldsymbol{Z}|\boldsymbol{\theta}).\notag
\end{align}
Furthermore, we have
\begin{align}
\ln p(\boldsymbol{X},\boldsymbol{Z}|\boldsymbol{\theta})&\!=\!\sum\nolimits_{n\in \cal N}\!\ln p(\boldsymbol{x}_n,\boldsymbol{z}_n|\boldsymbol{\theta})\notag\\\notag
&\!=\!\sum\nolimits_{n\in \cal N} \ln\prod\nolimits_{k\in \cal K}\![\pi_k p(\boldsymbol{x}_n|\boldsymbol{\theta}_k)]^{z_{nk}}\notag\\ \notag
&\!=\!\sum\nolimits_{n\in \cal N}\sum\nolimits_{k\in \cal K}\!z_{nk}[\ln \pi_k\!+\!\ln p(\boldsymbol{x}_n|\boldsymbol{\theta}_k)]\\
&\!=\!\sum\nolimits_{n\in \cal N}\sum\nolimits_{k\in \cal K}\!z_{nk}[\ln \pi_k\!+\!\ln p(\boldsymbol{r}_n|\boldsymbol{q}_n,\boldsymbol{\theta}_k)],\notag
\end{align}
where the second equality follows from \eqref{eq7}, and the last equality follows from \eqref{eq8}. As a result, the expectation of $\ln p(\boldsymbol{X},\boldsymbol{Z}|\boldsymbol{\theta})$ with respect to the posterior probabilities $p(\boldsymbol{Z}|\boldsymbol{X},\boldsymbol{\theta}^\text{old})$ can be rewritten as
\begin{align}
Q(\boldsymbol{\theta},\boldsymbol{\theta}^{\text{old}})&=\mathbb{E}_{p(\boldsymbol{Z}|\boldsymbol{X},\boldsymbol{\theta}^{\text{old}})}[\ln p(\boldsymbol{X},\boldsymbol{Z}|\boldsymbol{\theta})]\notag\\
&=\sum\nolimits_{n\in \cal N}\sum\nolimits_{k\in \cal K} \gamma_{nk}[\ln\pi_k+\ln p(\boldsymbol{r}_n|\boldsymbol{q}_n,\boldsymbol{\theta}_k)], \notag
\end{align}
where the identity $\gamma_{nk}= \mathbb{E}[z_{nk}]$ is used. Therefore, the optimization problem for the \emph{M-step} can be formulated as
\begin{align}
\textrm{(P3):} &\max_{\{\pi_k,\boldsymbol{\theta}_k\}}\sum_{n\in \cal N}\sum_{k\in \cal K} \gamma_{nk}[\ln\pi_k+\ln p(\boldsymbol{r}_n|\boldsymbol{q}_n,\boldsymbol{\theta}_k)]\notag \\
&~~~{\mathtt{s.t.}}\sum\nolimits_{k\in \cal K} \pi_k=1\notag\\
&~~~~~~~~0\le\pi_k\le1, \forall k\in \cal K. \notag
\end{align}
It is not difficult to see that problem \textrm{(P3)} can be decoupled into two independent sub-problems:
\begin{align}
\textrm{(P3.1):} &\max_{\{\pi_k\}}\sum\nolimits_{n\in \cal N}\sum\nolimits_{k\in \cal K} \gamma_{nk}\ln\pi_k \notag \\
&~{\mathtt{s.t.}}\sum\nolimits_{k\in \cal K} \pi_k=1\notag\\
&~~~~~~0\le\pi_k\le1, \forall k\in \cal K.\notag
\end{align}
\begin{align}
\textrm{(P3.2):} &\max_{\{\boldsymbol{\theta}_k\}}\sum\nolimits_{n\in \cal N}\sum\nolimits_{k\in \cal K} \gamma_{nk} \ln p(\boldsymbol{r}_n|\boldsymbol{q}_n,\boldsymbol{\theta}_k).\notag
\end{align}
\textrm{(P3.1)} is a convex optimization problem. With the standard Lagrangian method, its optimal solution can be obtained in closed form as $\pi^{*}_k={N_k}/{N}$, where $N_k=\sum_{n\in{\cal N}}\gamma_{nk}$, $\forall k\in{\cal K}$. On the other hand, the solution to \textrm{(P3.2)} depends on the actual conditional distribution $p(\boldsymbol{r}_n|\boldsymbol{q}_n,\boldsymbol{\theta}_k)$, i.e., the selected stochastic channel model. Based on the above results in the \emph{E-step} and \emph{M-step}, the general algorithm for EM-based CKM construction is summarized in Algorithm 2.
\begin{algorithm}[ht]
\small
\renewcommand{\algorithmicrequire}{\textbf{Input:}}
\renewcommand{\algorithmicensure}{\textbf{Output:}}
  \caption{General Algorithm for EM-based CKM Construction.}
  \begin{algorithmic}[1]
    \Require Given the stochastical channel model $p(\boldsymbol{r}|\boldsymbol{q},\boldsymbol{\theta})$;
    \State {\bf Initialization:} Choose initial parameters $\boldsymbol{\theta}^\text{old}$ and mixing coefficients $\pi_k^\text{old}$, $\forall k \in \cal K$;
    \State {\bf Repeat:}
        \begin{itemize}
            \item[1)] {\bf E-step}: Evaluate the {\it{responsibilities}} $\{\gamma_{nk}\}$ using \eqref{B1};
            \item[2)] {\bf M-step}: Update $\boldsymbol{\theta}^{\text{new}}$ by solving the optimization problem \textrm{(P3.2)}, and $\pi^{\text{new}}_k={N_k}/{N}$, with $N_k=\sum_{n\in{\cal N}}\gamma_{nk}$;
            \item[3)] $\boldsymbol{\theta}^{\text{old}}$ $\leftarrow$ $\boldsymbol{\theta}^{\text{new}}$, $\pi_k^{\text{old}}$ $\leftarrow$ $\pi_k^{\text{new}}$, $\forall k \in \cal K$;
        \end{itemize}
    \State {\bf Until} convergence or a maximum number of iterations is reached;
    \Ensure Set of modelling parameters $\boldsymbol{\theta}^{\text{new}}$, and {\it{responsibilities}} $\{\gamma_{nk}\}$.
  \end{algorithmic}
\end{algorithm}
\subsection{Special Case with CGM Construction}
In this subsection, we consider the the construction of a particular type of CKM, namely the CGM, for which the PDF of the channel gain given the user location is expressed as \eqref{eq2}. Eventually, the CGM can be directly constructed by using Algorithm 2. Here, by exploiting the specific structure of CGM, we obtain the closed-form expressions in \eqref{B1}, and obtain the optimal solution to (P3.2) in closed form, to reduce the construction complexity, as explained in detail as follows. First, $\gamma_{nk}$ in \eqref{B1} for the \emph{E-step} is given by the following closed-form expression:
\begin{align}
\gamma_{nk}=\frac{\pi_k^{\text{old}}N(r_n|\beta_k^{\text{old}}\!+\!10\alpha_k^{\text{old}}\log_{10}{\|\boldsymbol{q}_n\|},(\sigma_k^{\text{old}})^2))}{\sum\limits_{j\in \cal K} \pi_j^{\text{old}}
N(r_n|\beta_j^{\text{old}}\!+\!10\alpha_j^{\text{old}}\log_{10}{\|\boldsymbol{q}_n\|},(\sigma_j^{\text{old}})^2)}. \label{eq20}
\end{align}
Next, consider problem (P3.2), for which the log-likelihood function of the \emph{M-step} can be written as
\begin{align}
\ln p(r_n|\boldsymbol{q}_n,\boldsymbol{\theta}_k)\!=\! -\frac{1}{2}\ln(2\pi)-\frac{1}{2}\ln \sigma^2_k-\frac{(r_n\!-\!\beta_k\!-\!\alpha_kd_n)^2}{2\sigma^2_k}, \notag
\end{align}
where $d_n \triangleq 10\log_{10} \|\boldsymbol{q}_n\|$ is defined for convenience. By discarding constant terms, \textrm{(P3.2)} is equivalent to
\begin{align}
\min_{\{\alpha_k,\beta_k,\sigma^2_k\}}\sum_{n\in \cal N}\sum_{k\in \cal K}\! \gamma_{nk}\left[\ln\sigma^2_k\!+\!{(r_n\!-\!\beta_k\!-\!\alpha_kd_n)^2}/{\sigma^2_k}\right],\notag
\end{align}
which can be decoupled into $K$ independent sub-problems:
\begin{align}
\min_{\alpha_k,\beta_k,\sigma^2_k}\sum\nolimits_{n\in \cal N} \gamma_{nk}\left[\ln\sigma^2_k+{(r_n\!-\!\beta_k\!-\!\alpha_kd_n)^2}/{\sigma^2_k}\right].\label{eq23}
\end{align}
\begin{theorem}
The optimal solution to problem \eqref{eq23} is:
\begin{align}
\alpha_k&=\frac{\overline{(d_kr_k)}-\overline{d_k}\overline{r_k}}{\overline{d_k^2}-\overline{d_k}^2},~\beta_k=\frac{\overline{d_k^2}\overline{r_k}-\overline{d_k} \overline{(d_kr_k)}}{\overline{d_k^2}-\overline{d_k}^2},\label{eq25}\\
\sigma^2_k&={\sum\nolimits_{n\in \cal N}\gamma_{nk}(r_n-\beta_k-\alpha_kd_n)^2}/{N_k},\label{eq26}
\end{align}
where $\overline{d_k}=({\sum_{n \in \cal N}\gamma_{nk}d_n})/{N_k}$, $\overline{r_k}=({\sum_{n \in \cal N}\gamma_{nk}r_n})/{N_k}$,
$\overline{(d_kr_k)}=({\sum_{n \in \cal N}\gamma_{nk}d_nr_n})/{N_k}$, and
$\overline{d_k^2}=({\sum_{n \in \cal N}\gamma_{nk}d_n^2})/{N_k}$, with $N_k=\sum_{n\in{\cal N}}\gamma_{nk}$ $\forall k \in \cal K$.
\end{theorem}
\begin{IEEEproof}
Please refer to Appendix A.
\end{IEEEproof}
Based on the above derivations, the EM-based algorithm for the specific CGM construction is summarized in Algorithm 3.
\begin{algorithm}[ht]
\small
  \caption{EM-Based Algorithm for CGM Construction.}
  \begin{algorithmic}[1]
    \State {\bf Initialization:} Choose initial parameters $\boldsymbol{\theta}^\text{old}$ and mixing coefficients $\pi_k^\text{old}$, $\forall k \in \cal K$;
    \State {\bf Repeat:}
        \begin{itemize}
            \item[1)] {\bf E-step}: Evaluate the {\it{responsibilities}} $\{\gamma_{nk}\}$ using \eqref{eq20};
            \item[2)] {\bf M-step}: Update $\boldsymbol{\theta}^{\text{new}}$ using \eqref{eq25} and \eqref{eq26}, and update $\pi^{\text{new}}_k={N_k}/{N}$, with $N_k=\sum_{n\in{\cal N}}\gamma_{nk}$;
            \item[3)] $\boldsymbol{\theta}^{\text{old}}$ $\leftarrow$ $\boldsymbol{\theta}^{\text{new}}$, $\pi_k^{\text{old}}$ $\leftarrow$ $\pi_k^{\text{new}}$, $\forall k \in \cal K$;
        \end{itemize}
    \State {\bf Until} convergence or a maximum number of iterations is reached.
  \end{algorithmic}
\end{algorithm}
\subsection{Utilizing CKM for Channel Prediction}
Based on Algorithms 2 and 3, we obtain the modelling parameters $\boldsymbol{\theta}=\{{\mv \theta}_1,...,{\mv \theta}_K\}$, as well as the {\it{responsibilities}} $\{\gamma_{nk}\}$ for the measurement data $\boldsymbol{X}$. We are now ready to utilize such information to predict the channel knowledge $\mv r$ for any new location $\boldsymbol{q}$. To this end, we need to first determine which set of modelling parameters $\boldsymbol{\theta}_1,...,\boldsymbol{\theta}_K$ is most suitable for the new location $\boldsymbol{q}$. This can be achieved by determining the {\it{responsibilities}} $\{\gamma_k(\boldsymbol{q})\}$ using the IDW method. Specifically, let ${\cal M}$ denote the subset of $M$ user locations with training data that are nearest to $\boldsymbol{q}$. Then, $\gamma_k(\boldsymbol{q})$ is obtained as
$\gamma_k(\boldsymbol{q})=\sum\nolimits_{m \in \cal M} \omega_m\gamma_{mk}\label{gamma}$,
where $\omega_m={{d^{-1}_m(\boldsymbol{q})}}/{\sum_{j \in \cal M} {d^{-1}_j(\boldsymbol{q})}}$ is the weighting coefficient based on the IDW criterion, with $d_j(\boldsymbol{q})=\|\boldsymbol{q}_j-\boldsymbol{q}\|$, ${j\in {\cal M}}$. Then, the modelling parameter for location $\boldsymbol{q}$ is obtained as the one that maximizes $\gamma_k(\boldsymbol{q})$, i.e., $\boldsymbol{\theta}(\boldsymbol{q})=\boldsymbol{\theta}_{k^\star}$, where $k^\star=\arg \max\nolimits_{k \in \cal{K}} \gamma_k(\boldsymbol{q})$. As a result, the channel knowledge for location $\boldsymbol{q}$ can be predicted based on the PDF $p(\boldsymbol{r}|\boldsymbol{q}, \boldsymbol{\theta}(\boldsymbol{q}))$.
\section{Numerical Results}
\begin{figure}[ht]
\centerline{\includegraphics[width=6cm]{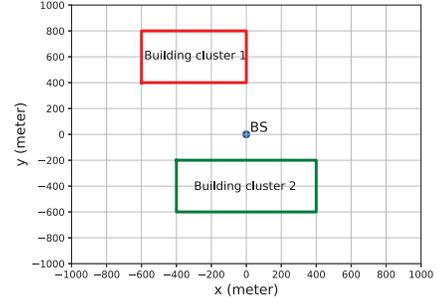}}\vspace{-1ex}
\caption{The layout for the considered wireless communication site.}
\label{fig}
\end{figure}

In this section, we present numerical results to validate the performance of our proposed algorithm. As shown in Fig. \ref{fig}, we consider a geographic area of size $2\times2~\text{km}^2$, and focus on the channel gains with a BS located at the center. We assume that there are two building clusters shown in Fig. \ref{fig}. Therefore, depending on the user locations, the direct line-of-sight (LoS) link may be blocked by one of the building cluster. Furthermore, for those indoor users located in the building cluster area, additional penetration loss is incurred. As a result, depending on the user locations, the groundtruth channel gains are generated based on 5 user groups: LoS users that have direct LoS link with the BS; NLoS1 and NLoS2 users whose LoS links are blocked by building clusters 1 and 2, respectively; and Indoor1 and Indoor2 users that are located in building clusters 1 and 2, respectively. The corresponding modelling parameters of each user group are given in Table \ref{tab1}. Furthermore, the initial parameters of Algorithm 3 are set as $\pi_k^\text{old}=1/K,\forall k\in{\cal K}$, and ${\mv \theta}^\text{old}$ are randomly generated as $\alpha_k\in[2,5]$, $\beta_k\in[30,140]$, and $\sigma^2_k\in[6,15],\forall k\in{\cal K}$.
Unless otherwise stated, the number of data points used for training is $N=2000$.
\begin{table}[ht]
\caption {Groundtruth Modelling Parameters.}\vspace{-1ex}
\centering
\begin{tabular}{|c|c|c|c|c|}
\hline
\multicolumn{1}{|c|}{{User group} }&$\alpha$ & $\beta$& $\sigma^2$\\
\hline
\multirow{5}*{}LoS&2.2&30&6.25\\
\cline{1-4}
NLoS1&2.6&55&10.24\\
\cline{1-4}
NLoS2&3.1&80&10.24\\
\cline{1-4}
Indoor1&3.6&105&7.84\\
\cline{1-4}
Indoor2&4.1&130&7.84\\
\hline
\end{tabular}
\label{tab1}	
\end{table}
\begin{figure}[ht]
\centerline{\includegraphics[width=6cm]{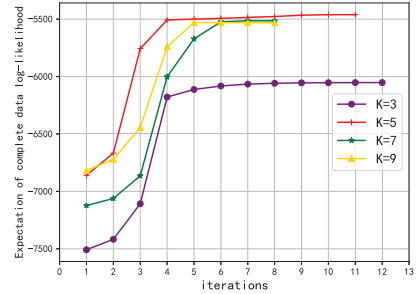}}\vspace{-1ex}
\caption{Convergence of Algorithm 3 for CGM construction.}
\label{fig2}
\end{figure}
\begin{figure*}[ht]
\centering
\hspace{30pt}
\subfigure[Groundtruth data points.]{
\includegraphics[width=5.9cm]{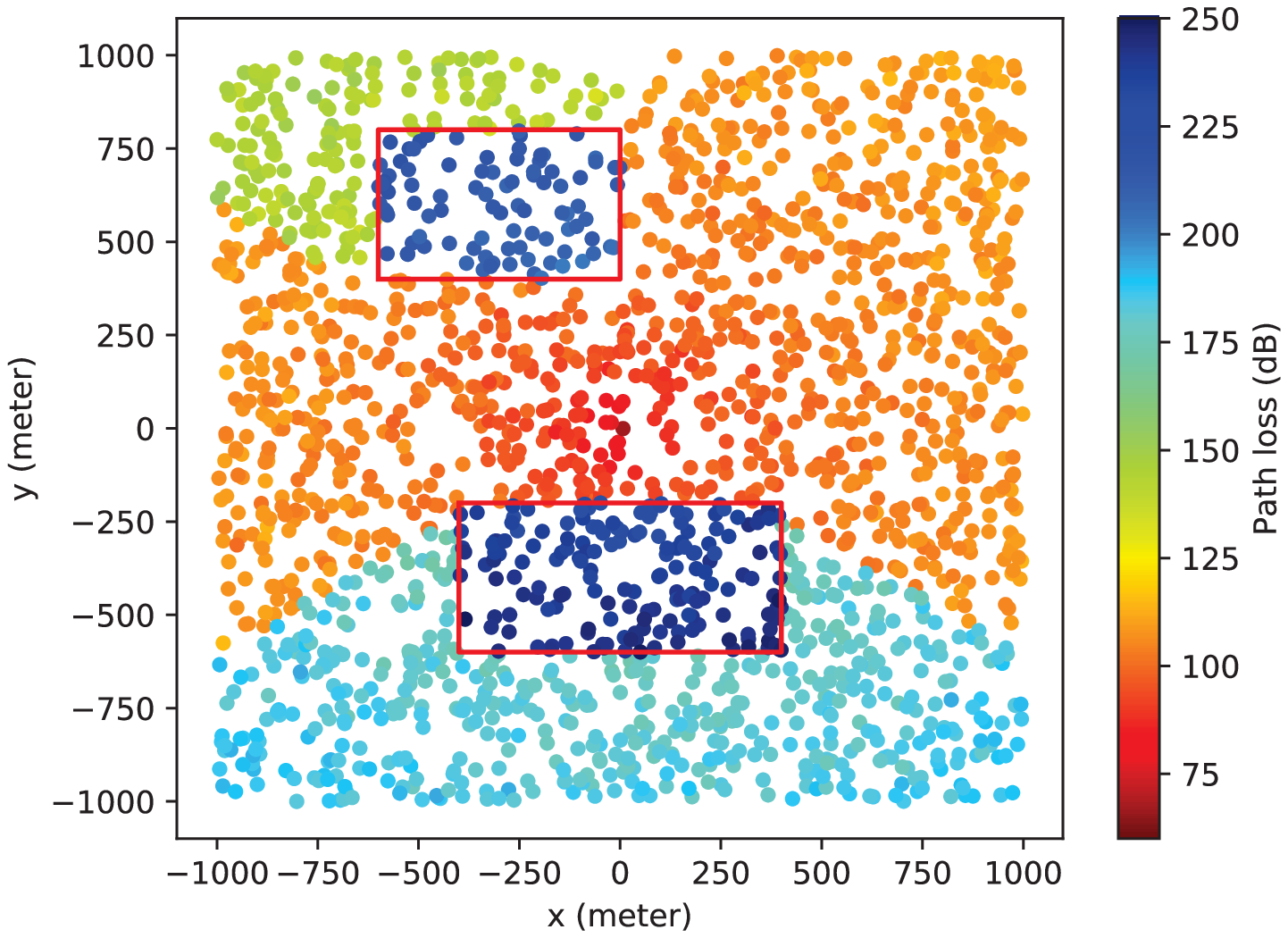}
}\hspace{-38pt} 
\quad
\subfigure[EM algorithm, $K=5$.]{
\includegraphics[width=5.9cm]{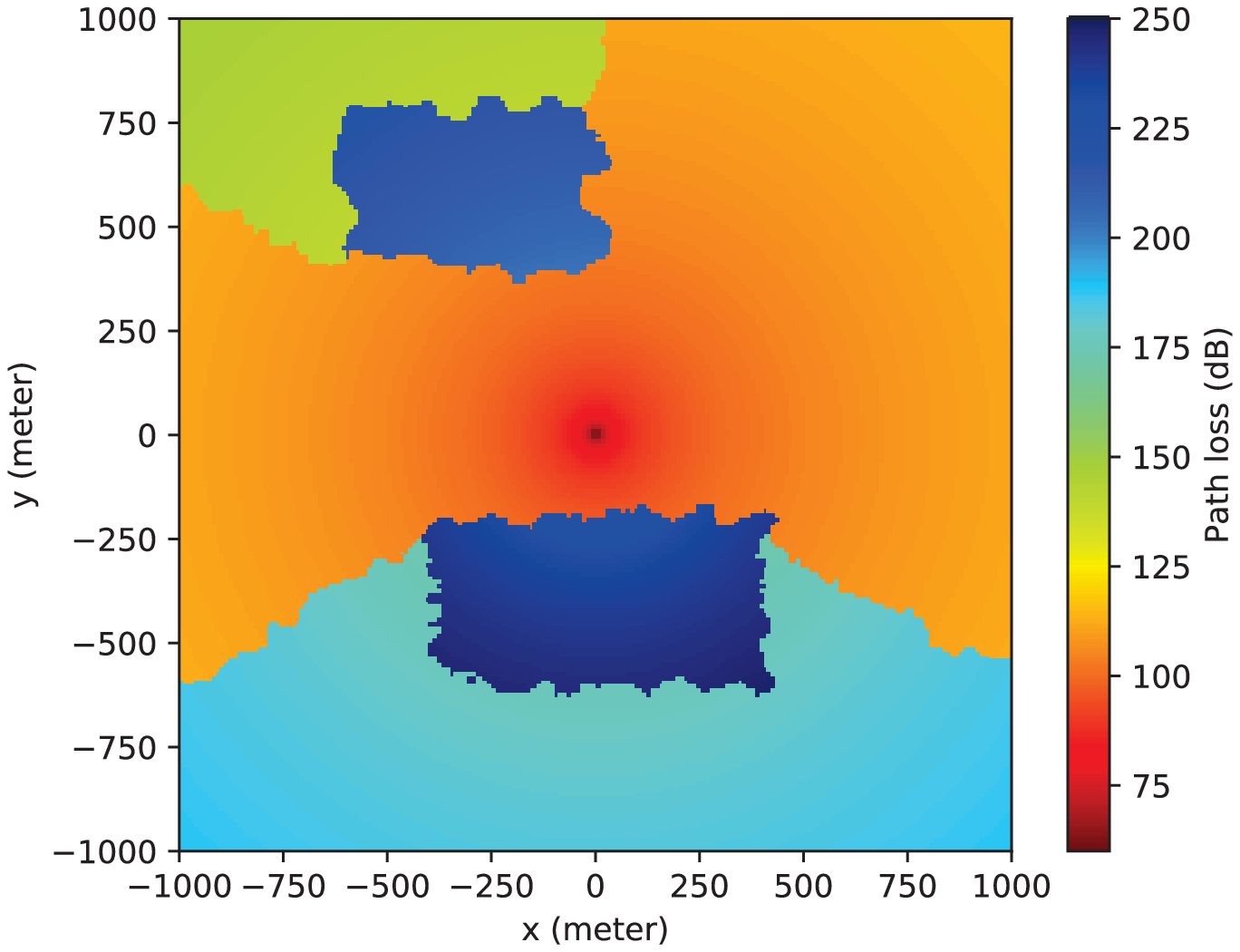}
}\hspace{-38pt}
\quad
\subfigure[Single model curve fitting.]{
\includegraphics[width=5.9cm]{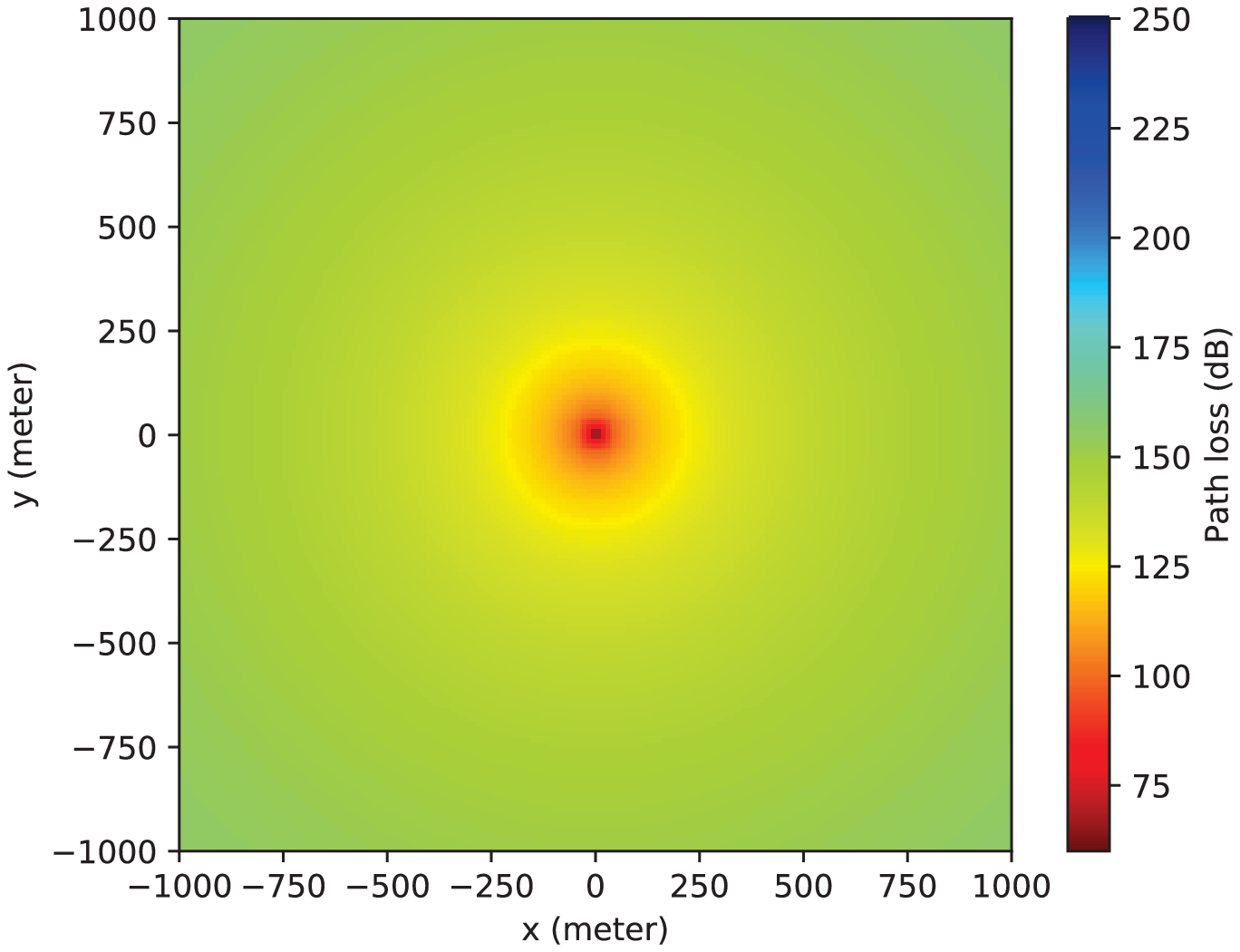}
}\vspace{-1ex}
\caption{Comparison of the constructed CGM based on the proposed EM algorithm and single model curve fitting method.}
\label{fig1}
\end{figure*}

First, to show the convergence of Algorithm 3, Fig. \ref{fig2} plots the expectation of the {\it{complete-data}} log likelihood $Q({\mv \theta}^\text{new}, {\mv \theta}^\text{old})$ versus the iteration number of each EM cycle for different number of assumed modelling components $K$. It is observed that for all the $K$ values considered, Algorithm 3 results in monotonically non-decreasing log likelihood values, which guarantees the convergence. Furthermore, when the assumed number of components matches with the groundtruth value, i.e., $K=5$, a fastest convergence is observed. On the other hand, with $K\geq 5$, Algorithm 3 is still able to converge to roughly the same components number matches as $K=5$, though at slightly slower rate, while that for $K=3$ leads to poor performance since the number of assumed mixing components is smaller than the groundtruth.

Next, we evaluate the quality of the constructed CGM based on Algorithm 3. For comparison, we also consider a benchmark scheme based on the single-model curve fitting method, i.e., with $K=1$. In this case, no latent variable is involved since all available data points will be fitted to one single set of parameters. Fig. \ref{fig1}(a) shows the groundtruth data points, and Fig. \ref{fig1}(b) and Fig. \ref{fig1}(c) plot the constructed CGMs with $K=5$ and $K=1$, respectively, by using the IDW method with $M=3$.
It is observed that compared to the single-model curve fitting method, the proposed EM-based algorithm results in more accurate CGM. In fact, it is observed from Fig.~\ref{fig1}(c) that the single-model curve fitting method leads to concentric contours of channel gain, which is far from the reality as shown in Fig.~\ref{fig1}(a). This is due to the fact that the conventional single-model curve fitting method cannot distinguish the characteristics of different data points at different sub-areas. This issue can be addressed by our proposed EM-based algorithm with mixed channel models, as shown in Fig. \ref{fig1}(b).

\begin{figure}[ht]
\centerline{\includegraphics[width=6cm]{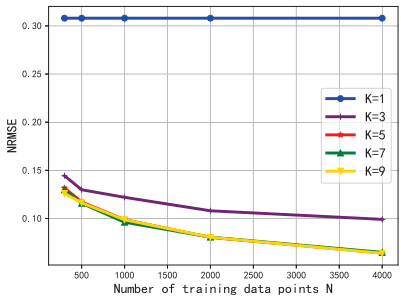}}\vspace{-1ex}
\caption{NRMSE of the predicted channel gain versus the number of training data points $N$.}
\label{fig3}
\end{figure}

To evaluate the impact of the number of training data points $N$, Fig. \ref{fig3} plots the normalized root mean square error (NRMSE) of the predicted channel gains versus $N$. The testing set consists of 1000 data points.
It is observed from Fig. \ref{fig3} that as $N$ increases, the constructed CGM by the proposed EM algorithm with $K>1$ has better fitting quality. By contrast, regardless of $N$, the single model curve fitting method with $K=1$ has poor performance, which is due to the fact that the number of tunable modelling parameters is too small to accurately fit the complex environment. This demonstrates the effectiveness of our proposed EM algorithm for accurate CKM construction in complex environments.
\section{Conclusion}
In this paper, we proposed a novel EM-based CKM construction method towards environment-aware communications, by utilizing both the available measurement data points and the expert knowledge with well-established statistic channel models. The key idea is to partition the available data points into different groups, where each group shares the same modelling parameter values that are to be determined. We propose to use the classic EM algorithm to determine the modelling parameters by solving an equivalent maximum likelihood estimation problem with latent variables, and then extend the algorithm for constructing the specific CGM. Numerical results demonstrated the effectiveness of the proposed algorithm as compared to the benchmark curve-fitting scheme with one single model.
How to extend the results to other types of channel knowledge and the efficient utilizations of the constructed CKM are interesting directions worth pursuing in future research.
\appendices
\section{Proof of Theorem 1} \label{App}
Problem \eqref{eq23} can be solved by setting the partial derivatives with respect to the optimization variables to be zero. For convenience, let $f$ denote the cost function of problem \eqref{eq23}. Therefore, we have
\begin{align}
\frac{\partial f}{\partial \alpha_k}&=-\frac{2}{\sigma^2_k}\sum\nolimits_{n\in \cal N} \gamma_{nk}(r_n-\beta_k-\alpha_kd_n)d_n=0\label{B4},\\
\frac{\partial f}{\partial
\beta_k}&=-\frac{2}{\sigma^2_k}\sum\nolimits_{n\in \cal N} \gamma_{nk}(r_n-\beta_k-\alpha_kd_n)=0. \label{B5}
\end{align}
Then, \eqref{B4} and \eqref{B5} can be arranged as
\begin{align}
\beta_k\sum_{n\in \cal N} \gamma_{nk}d_n+\alpha_k\sum\limits_{n\in \cal N}\gamma_{nk}d_n^2&=\sum_{n\in \cal N} \gamma_{nk}r_nd_n,\notag\\
\beta_k\sum\nolimits_{n\in \cal N} \gamma_{nk}+\alpha_k\sum\nolimits_{n\in \cal N}\gamma_{nk}d_n&=\sum\nolimits_{n\in \cal N}\gamma_{nk}r_n. \notag
\end{align}
By solving the linear system equations with two unknowns and two equations, we get \eqref{eq25}. Similarly, by setting the partial derivative of $f$ with respect to $\sigma_k^2$ to zero, we get \eqref{eq26}. This thus completes the proof.

\end{document}